\documentclass[journal,twoside,web]{ieeecolor}
\usepackage{lcsys}
\usepackage{mathtools}
\usepackage{amssymb}
 \usepackage{booktabs} 
\newcommand{\rvline}{\hspace*{-\arraycolsep}\vline\hspace*{-\arraycolsep}}
\usepackage{graphicx}
\usepackage{algorithm}
\usepackage{algpseudocode}
\usepackage[normalem]{ulem}
\usepackage[dvipsnames]{xcolor}
\newtheorem{thm}{Theorem}[section]
\newtheorem{lemma}{Lemma}[section]

\newtheorem{rem}{Remark}[section]

\renewcommand{\vec}{\mathrm{vec}}
\newcommand{\vech}{\mathrm{vech}}

\title{\LARGE \bf State Estimation for Linear Systems with Quadratic Outputs}

	\author{Soulaimane~Berkane, \IEEEmembership{Senior Member, IEEE}, Dionysis~Theodosis, Tarek Hamel, \IEEEmembership{Fellow, IEEE}\\ Dimos~V.~Dimarogonas, \IEEEmembership{Fellow, IEEE} %
		\thanks{S. Berkane is with the D\'epartement   d'informatique   et   d'ing\'enierie, Universit\'e   du   Qu\'ebec   en   Outaouis,   Gatineau, Qu\'ebec,   Canada (e-mail: \texttt{soulaimane.berkane@uqo.ca}). S. Berkane is also with the Department of Electrical Engineering, Lakehead University, Ontario, Canada.  D. Theodosis is with the Dynamic Systems and Simulation Laboratory,  
		Technical University of Crete, Greece (e-mail: \texttt{dtheodosis@dssl.tuc.gr}). D. V. Dimarogonas is with the Division of Decision and Control Systems, School of Electrical Engineering and Computer Science, KTH Royal Institute of Technology, Sweden (e-mail: \texttt{dimos@kth.se}). T. Hamel is with I3S-CNRS, University C\^ote d'Azur and Institut universitaire de France, France (e-mail: \texttt{thamel@i3s.unice.fr}).}%
		\thanks{* This research work is supported in part by NSERC-DG RGPIN-2020-04759, Fonds de recherche du Qu\'ebec (FRQ),  the Swedish Research Council (VR), and the Knut och Alice Wallenberg foundation (KAW).}} %

\begin{document}
\maketitle
\thispagestyle{empty}
\begin{abstract}
This letter deals with the problem of state estimation for a class of systems involving linear dynamics with multiple quadratic output measurements. We propose a systematic approach to immerse the original system into a linear time-varying (LTV) system of a higher dimension.  The methodology extends the original system by incorporating a minimum number of auxiliary states, ensuring that the resulting extended system exhibits both linear dynamics and linear output. Consequently, any Kalman-type observer can showcase global state estimation, provided the system is uniformly observable. 
\end{abstract}


\section{Introduction}\label{sec:intro}
\subsection{Motivation}

\IEEEPARstart{S}{tate} estimation is a  central problem in control theory, with widespread applications in various engineering fields. It consists of designing a software-implemented dynamical system, known as {\it state observer}, that allows the determination of the system's internal state from measurements of its inputs and outputs. Several approaches have been proposed to address this problem; please refer to \cite{B22}, \cite{B07}, \cite{GK01}, \cite{T90}, and related references therein. 

In this paper,  we consider systems with linear dynamics and quadratic output measurements, \textit{i.e.,} systems of the form
\begin{subequations}\label{sys:M}
\begin{align}
\dot{x}=&Ax+Bu,\label{sys:M:dyn}\\
y_h=&\frac{1}{2}x^\top C_hx+d_h^\top x ,\,\,h=1,\ldots,q,\label{sys:M:out}
\end{align}  
\end{subequations} 
where $x\in\mathbb{R}^n$ is the state, $u\in\mathbb{R}^{p}$ is the input,  and $y_1, \cdots, y_q\in\mathbb{R}$ are scalar outputs. The system matrices  $A\in\mathbb{R}^{n\times n}$, $B\in\mathbb{R}^{n\times p}, C_h\in\mathbb{R}^{n\times n},$ and $d_h\in\mathbb{R}^{n\times 1}$ are constant and, without any loss of generality, the matrices $C_h$ are assumed to be symmetric.  
Note that system \eqref{sys:M} belongs to a specific class of nonlinear systems, and its observability properties are affected by the input $u$. In particular, for the trivial system $\dot{x}=0$, $y=x^2$, distinguishing between the initial conditions $x_0$ and $-x_0$ becomes impossible solely based on the provided output measurements.

Systems exhibiting linear dynamics and quadratic outputs are prevalent in diverse control and estimation scenarios. Control of systems with quadratic outputs has been dealt with in \cite{montenbruck2017linear} where the authors' partial motivation stemmed from mechanical systems, in which one wishes to regulate specific
energies, {\it e.g.,} \cite{van1986stabilization,bloch2000controlled}. Range measurements, often used in robot localization problems, can be written as quadratic outputs. Range-based state estimation techniques have been proposed, {\it e.g.,} in \cite{A21,BSO11,PI17,IDP16,HS17}, mainly dealing with single and double integrator systems. In \cite{F19}, if quadratic functions approximate known terrain maps,  terrain-aided navigation can be achieved by designing an observer for the vehicle's dynamics (single- or double-integrator systems) under quadratic output measurements.
 
\subsection{Literature Review}
For the design of an observer, one needs an appropriate notion of observability, namely, the ability to deduce the initial state vector by using the input and corresponding output information across a given time period. Typically, when studying the observability of a nonlinear system, it is commonly viewed as a local problem. The {\it local weak observability} of such a system can be determined using the standard observability rank condition \cite[Theorem 3.1]{hermann1977nonlinear}. However, this condition alone is insufficient for designing an observer, as it heavily relies on the system's input. In such cases, the design process is limited to specific classes of inputs, namely regular or persistently exciting inputs. More information on these input classes can be found in \cite{B07}, \cite{besanccon1996observer}, \cite{bornard1989regularly}, \cite{GK01}, and related sources.

To design observers for nonlinear systems, a well-known technique is the {\it immersion approach}, where the nonlinear system is transformed into a state-affine system with a higher dimension. This methodology has a long history of development. In \cite{FK83}, a necessary and sufficient condition for immersion based on the system's observation space was presented. Another approach, as discussed in \cite{BS04} and \cite{J03}, utilized the solutions of a partial differential equation for the immersion process. Additionally, in \cite{BT07}, an immersion-based technique was introduced for a broad range of nonlinear systems (that are rank-observable), employing a high-gain design strategy.

For systems with quadratic outputs, such as the one represented in \eqref{sys:M}, approaches outlined in \cite{B07, ciccarella1993luenberger, GK01, gauthier1992simple} demonstrate the feasibility of transforming the system into a canonical form by leveraging a local change of coordinates for suitable inputs. It's important to note, however, that these methods yield local results. It should be noted that algebraic conditions for the observability of such systems were also proposed in \cite{D71}, which, however, do not guarantee state reconstruction.
 
\subsection{Contributions}

To estimate the state of system \eqref{sys:M}, a systematic approach is presented that transforms the original system into a new bilinear system of higher dimension of the form \begin{subequations}\label{sys:noC:ext}
\begin{align}
\dot{z}&=\mathcal{A}(u )z+\mathcal{B}u\label{sys:noC:ext:s},\\
y&=\mathcal{C} z,\label{sys:noC:ext:out}
\end{align}
\end{subequations}
where $z\in\mathbb{R}^{m+n}$ for certain integer $m\ge1$, $y:=[y_1\;\cdots y_q]^\top$, with $\mathcal{A}(u)$, $\mathcal{B}$, and $\mathcal{C}$ of appropriate dimensions. Notice that \eqref{sys:noC:ext:s} is a bilinear system since the matrix function $\mathcal{A}(u)$ depends explicitly on the input $u$ and can be considered as an LTV system once an input time-function $u(t)$ is fixed.

We first show that the observation space is finite-dimensional with a dimension equal to a constant $m$ less than or equal to $n(n+1)/2$ (dimension of the space of symmetric matrices). Then, we provide a simple algorithm to extract this constant $m$ and the corresponding {\it basis} for the observation space. Then, by extending the original system with $m$ additional variables, the new dynamics of the resulting system are in the form of \eqref{sys:noC:ext}, where we explicitly identify the involved matrices. This work generalizes our results in \cite{TBD20} where we considered systems with {\it single output} for which the zero-response (output under zero input) is {\it polynomial}. In contrast, the current work immerses the general {\it multiple output system} into an LTV system with minimal auxiliary states. There is no restriction on the number of outputs nor on the time behavior of the output.

Compared to those above (local) nonlinear estimation techniques, the proposed approach allows the design of an observer to ensure global exponential convergence (to zero) of the state estimation errors using simple linear systems tools, and linear Kalman-type observers, see \cite{B07}, \cite{Kalman1961new}, \cite{Rugh96}.

\section{Notation}\label{sec:prel}
Throughout this paper, we adopt the following notation. $\mathbb{N}$ and $\mathbb{R}$ denote, respectively, the sets of natural and real numbers.
For a given vector or matrix $(\cdot)\in{\mathbb R}^{n\times m}$, $(\cdot)^{\top}$ denotes its transpose.
We denote by $I_{n}$ the $n\times n$ identity matrix. By $0$ we denote each of the following: the scalar zero, the zero vector, or the zero matrix. Depending on the context, the dimensions of $0$ will be clear unless otherwise specified as $0_{n\times m}$.  Let $\mathbb{M}_n\subset\mathbb{R}^{n\times n}$ denote the space of real $n\times n$ symmetric matrices. For a given map $f$, we denote by $f^{[n]}$ the $n$-th iterate of $f$ (functional power) such that $f^{[n]}:=f\circ f^{[n-1]}$ and $f^{[0]}$ is the identity map. For any matrix $A\in\mathbb{R}^{n\times n}$, we define the Lyapunov operator $\mathbf{L}_A:\mathbb{M}_n\to\mathbb{M}_n$  by $\mathbf{L}_A(X):=XA+A^\top X$. From the previous definition, we also have $\mathbf{L}_A^{[k]}(C)=\mathbf{L}_A\circ \mathbf{L}_A^{[k-1]}(C)$, $\mathbf{L}_A^{[0]}(C)=C$.  A {\it permutation matrix} is an $n\times n$ square matrix that has exactly one entry equals $1$ in each row and each column and $0$s elsewhere. Note that any permutation matrix $P$ is orthogonal, {\it i.e.,} $P^{-1}=P^\top$. Given a linear map $T:V\to W$  between two vector spaces $V$ and $W$, we define the kernel of  $T$ by $\ker(T)=\{v\in V: T(v)=0\in W\}$.

For a given matrix $A\in\mathbb{R}^{n\times m}$, $\vec(A)$ denotes the {\it vectorization} of $A$ which is the $nm\times 1$ column vector obtained by stacking the columns of the matrix $A$ on top of one another.  For a symmetric matrix $A\in\mathbb{M}_n$, the {\it half-vectorization} of $A$, denoted by $\vech(A)$, is the $n(n + 1)/2\times 1$ column vector obtained by vectorizing only the lower triangular part of $A$. The {\it duplication matrix} (see \cite{magnus2019matrix}) is the unique $n^2\times n(n + 1)/2$ matrix, denoted by $D_n$, which, for any symmetric matrix $A\in\mathbb{M}_n$, transforms $\vech(A)$ into $\vec(A)$, \textit{i.e.,}
\begin{align}
\label{eq:vec-vech}
    D_n\vech(A)=\vec(A).
\end{align}
Note that $D_n$ is full column rank and, hence, its Moore-Penrose inverse is given by
$
    D_n^+=(D_n^\top D_n)^{-1}D_n^\top.
$
For any two matrices $A\in\mathbb{R}^{n\times m}$ and $B\in\mathbb{R}^{p\times q}$, we use $A\otimes B\in\mathbb{R}^{np\times mq}$ to denote the Kronecker product of $A$ and $B$. The following are some useful properties of the (associative) Kronecker product \cite{brewer1978kronecker}:
\begin{align}
\label{eq:KP1}
    \vec(ACB)=(B^\top\otimes A)\vec(C),\\
\label{eq:KP2}
    (A\otimes B)(C\otimes D)=(AC)\otimes(BD).
\end{align}
Moreover, for any vectors $u\in\mathbb{R}^m, v\in\mathbb{R}^q,$ and for any matrix $A\in\mathbb{R}^{n\times m}$, straightforward applications of \eqref{eq:KP2} results in the following identities:
\begin{align}
\label{eq:KP3}
    (Au)\otimes v=(A\otimes I_q)(u\otimes v),\\
\label{eq:KP4}
    v\otimes (Au)=(I_q\otimes A)(v\otimes u),\\
\label{eq:KP5}
    u\otimes v=(I_m\otimes v)u=(u\otimes I_q)v.
\end{align}
We also provide the following useful identity, whose proof follows directly from \cite[Theorem 9]{magnus2019matrix} and \cite[Theorem 12]{magnus2019matrix}:
\begin{align}
    \label{eq:KP9}
    (D_mD_m^+)^\top (u\otimes u)&=u\otimes u.
\end{align}
Finally, for any two square matrices $A\in\mathbb{R}^{n\times n}$ and $B\in\mathbb{R}^{m\times m}$, we define their {\it Kronecker sum} by
\begin{align}
\label{eq:KP6}
    A\oplus B:=A\otimes I_m+I_n\otimes B.
\end{align}
In the following useful lemma, we show that the composition of the half-vectorization with the Lyapunov operator is a linear map on $\mathbb{R}^{n(n+1)/2}$.
\begin{lemma}\label{lemma:vech-LA}
For any $A\in\mathbb{R}^{n\times n}$ and $X\in\mathbb{M}_n$, one has
\begin{equation}
    \vech(\mathbf{L}_A(X))=D_n^+(A\oplus A)^\top D_n\vech(X).
\end{equation}
\end{lemma}
The proof of Lemma \ref{lemma:vech-LA} can be found in the Appendix.
\section{Immersion into an LTV system}\label{sec:augmentation}
\subsection{Single-Output Systems }\label{ssec:Mot}
To motivate our general methodology, we first consider single-output systems.  First, note that any \textit{quadratic form}, along the trajectories of \eqref{sys:M:dyn}, satisfies
\begin{align}\label{eq:quadratic_form_derivative}
        \frac{1}{2}\frac{d}{dt}(x^\top Qx)=\frac{1}{2}x^\top\mathbf{L}_A(Q)x+(QBu)^\top x,
\end{align}
for some symmetric matrix $Q\in\mathbb{M}_n$. Therefore, and for general linear systems, the time derivative of any quadratic form is a (time-varying) \textit{quadratic function}. Also, \eqref{eq:quadratic_form_derivative} shows that, for any quadratic form, successive Lie derivatives along the vector field of a linear system are quadratic forms, all characterized by a symmetric matrix $\mathbf{L}_A^{[k]}(Q)$ for some $k\in\mathbb{N}$. Note that, thanks to the space of symmetric matrices $\mathbb{M}_n$ being finite-dimensional, the so-called {\it finiteness criterion} of the observation space \cite{FK83} is, therefore, satisfied. This vector subspace of $\mathbb{M}_n$ can have a lower dimension compared to $\mathbb{M}_n$. This is shown in the following result.
\begin{lemma}\label{lemma:Qm}
    For any $Q\in\mathbb{M}_n$, there exists $m\leq n(n+1)/2$ such that $\mathrm{span}\{\mathbf{L}_A^{[0]}(Q), \mathbf{L}_A^{[1]}(Q),\ldots,\mathbf{L}_A^{[m-1]}(Q)\}$ is $\mathbf{L}_A-$invariant.
\end{lemma}
The proof of Lemma \ref{lemma:Qm} can be found in the Appendix.

Now, consider system \eqref{sys:M} with single output (\textit{i.e.,} $q=1$, index $h$ is hence ignored subsequently). Clearly the output map \eqref{sys:M:out} is {\it representative} (see \cite[Definition 7]{FK83}) since its observation space is finite-dimensional by Lemma \ref{lemma:Qm}. By \cite[Theorem 1]{FK83}, system \eqref{sys:M} with single output is representable as a subsystem of an affine system.  To construct such a system, we define the following auxiliary state variables
\begin{equation}\label{zi:state}
{\begin{aligned}
	\xi_k:=&\frac{1}{2}x^{\top}\mathbf{L}_A^{[k]}(C)x  ,\,\,\, k=0,1,\cdots,(m-1),
\end{aligned}}
\end{equation}
with $m\leq n(n+1)/2$ being the {\it smallest index} satisfying Lemma \ref{lemma:Qm}. Now, in view of \eqref{eq:quadratic_form_derivative}, one has
\begin{equation}\label{eq:dzi1}
    {\begin{aligned}
        \dot{\xi}_k=&\xi_{k+1}+ (Bu)^\top\mathbf{L}_A^{[k]}(C) x,\; k=0,\ldots,(m-2).
    \end{aligned}}
\end{equation}
Moreover, in view of \eqref{eq:quadratic_form_derivative} and Lemma \ref{lemma:Qm}, there exist scalars $\alpha_1,\cdots,\alpha_{m-1}$ such that
\begin{equation}
\begin{aligned}
        \dot{\xi}_{m-1}=\sum_{k=0}^{m-1}\alpha_k \xi_k+ (Bu)^\top\mathbf{L}_A^{[m-1]}(C) x.
\end{aligned}
\end{equation}
Define now the extended state as follows
\begin{equation}\label{ext:state:1}
z:=\left[\begin{matrix}
\xi_{m-1}&\cdots&\xi_1&\xi_0&x^\top
\end{matrix}\right]^{\top}\in\mathbb{R}^{m+n}.
\end{equation}
In view of \eqref{sys:M} and \eqref{eq:dzi1}, the dynamics of the new variable $z\in\mathbb{R}^{m+n}$ are given by the LTV system \eqref{sys:noC:ext},
where the matrices $\mathcal{A}(u)\in\mathbb{R}^{(m+n)\times(m+n)}$, $\mathcal{B}\in\mathbb{R}^{(m+n)\times p}$ and $\mathcal{C}\in\mathbb{R}^{1\times(m+n)}$ are defined by
\begin{align}\nonumber
&\mathcal{A}(u):=\\
&\begin{bmatrix}
\alpha_{m-1}&\alpha_{m-2}&\ldots&\alpha_1 &\alpha_0&\rvline&(Bu)^\top\mathbf{L}_A^{[m-1]}(C)\\
1&0&\ldots &0&0&\rvline&(Bu)^\top\mathbf{L}_A^{[m-2]}(C)\\
\vdots&\vdots&\vdots&\ddots &\vdots&\rvline&\vdots\\
0&0&\ldots &1&0&\rvline&(Bu)^\top\mathbf{L}_A^{[0]}(C)\\
\cmidrule(lr){1-7}
0&0&\cdots&0&0&\rvline&A
\end{bmatrix},
\label{matrix:Au:ext}\\
\mathcal{B}&:=\left[\begin{matrix}
0&0&\cdots&\rvline&B^\top
\end{matrix}\right]^\top,\\
\mathcal{C}&:=\left[\begin{matrix}  \begin{matrix}
0&0&\cdots&0&1
\end{matrix}
&\rvline&d^\top\end{matrix}\right]\label{matrix:Cu:ext}.
\end{align}
The following theorem summarizes the results above.
\begin{thm}\label{theorem:single-ouput}
Consider system \eqref{sys:M}  along with matrices $A, B, C, d$ of appropriate dimensions.  If the system involves only a single output ($q=1$), then there exists $m\leq n(n+1)/2$ and coefficients $\alpha_0,\cdots,\alpha_{m-1}$ satisfying Lemma \ref{lemma:Qm} such that system \eqref{sys:M} can be immersed into the LTV system \eqref{sys:noC:ext} with matrices $\mathcal{A}(u)$, $\mathcal{B}$, and $\mathcal{C}$ given by \eqref{matrix:Au:ext}-\eqref{matrix:Cu:ext}.

\end{thm}

This result shows that the maximum number of additional states needed to bring the single-output dynamical system  \eqref{sys:M} to the LTV form cannot exceed $n(n+1)/2$. This latter property is a consequence of the fact that the vector space $\mathbb{M}_{n}$ (space of symmetric matrices) has dimension $\dim(\mathbb{M}_{n})=n(n+1)/2$. Therefore, any quadratic form on $\mathbb{R}^n$ can be expressed as a linear combination of $n(n+1)/2$ (elementary) quadratic forms. However, it is important to pick up the smallest $m\leq n(n+1)/2$ satisfying Lemma \ref{lemma:Qm} to avoid adding unnecessary auxiliary states, which might introduce additional observability restrictions. Having fewer auxiliary states in the obtained LTV system leads to weaker observability conditions for the LTV system. The following are some illustrative examples where $m$ usually takes smaller values compared to $n(n+1)/2$.

\textit{Example\;1}\label{example1}: Consider the double-integrator system $\ddot x=u\in\mathbb{R}^n$ with $y=0.5\|x\|^2$. We have $$\mathbf{L}_A^{[1]}(C)=\begin{bmatrix}
0&I_n\\I_n&0
\end{bmatrix},\;\;\mathbf{L}_A^{[2]}(C)=\mathbf{L}_A(\mathbf{L}_A^{[1]}(C))=\begin{bmatrix}
0&0\\ 0&2I_n
\end{bmatrix} $$ and finally, $\mathbf{L}_A^{[3]}(C)=0$. Hence, $m=3$ satisfies Lemma \ref{lemma:Qm} and Theorem \ref{theorem:single-ouput}, $\forall n\in\mathbb{N}$.

\textit{Example\;2}\label{example3}: Consider the system $\dot x_1=x_2, \dot x_2=x_1+2x_2+u\in\mathbb{R}$ with $y=0.5x_1^2+0.5x_2^2$. We have
$$\mathbf{L}_A^{[1]}(C)=\begin{bmatrix}
 0&2\\2&4   
\end{bmatrix},\;\;\mathbf{L}_A^{[2]}(C)=\begin{bmatrix}
    4&8\\8&20
\end{bmatrix}$$
It is observed that $\mathbf{L}_A^{[2]}(C)=4(\mathbf{L}_A^{[0]}(C)+\mathbf{L}_A^{[1]}(C))$ (recall that $\mathbf{L}_A^{[0]}(C)=C=I_2$) and, hence, $m=2$ satisfies Lemma \ref{lemma:Qm}. In this case, the number of auxiliary states needed is $m=2$, which is strictly less than $n(n+1)/2=3$.

\subsection{Multi-Output Systems}\label{ssec:M}
We now consider systems with multiple quadratic outputs of the form  \eqref{sys:M}. The development hereafter is based on the observation that the outputs \eqref{sys:M:out} can be expressed as follows:
\begin{equation}
\begin{aligned}
    y_h&=\frac{1}{2}x^\top C_hx+d_h^\top x\\
    &\stackrel{\eqref{eq:KP1}}{=}\frac{1}{2}\vec(C_h)^\top (x\otimes x)+d_h^\top x\\
    &\stackrel{\eqref{eq:vec-vech}}{=}\frac{1}{2}\vech(C_h)^\top D_n^\top (x\otimes x)+d_h^\top x.
\end{aligned}
\end{equation}
If we define the vector $x^{[2]}:=D_n^\top (x\otimes x)$, the output vector $y:=[y_1\cdots y_q]^\top$ can be written as
\begin{equation}
    y=\frac{1}{2}\begin{bmatrix}
    \vech(C_1)^\top \\
    \vdots\\
    \vech(C_q)^\top 
    \end{bmatrix}x^{[2]}+\begin{bmatrix}
    d_1^\top \\
    \vdots\\
    d_q^\top 
    \end{bmatrix}x=:\bar Cx^{[2]}+Dx.
\end{equation}
Before we proceed, some remarks are in order. The vector $x^{[2]}$ contains  $n(n+1)/2$ linearly independent $2-$degree terms of the form $\alpha x_ix_j$ for $1\leq i\leq j\leq n$. These terms are proportional to $x_ix_j=x^\top E_{i,j} x$ where $\{E_{i,j}\}_{1\leq i\leq j \leq n}$, is the canonical basis of the vector space $\mathbb{M}_{n}$. The number of these terms is $n(n+1)/2$, equal to the dimension of the space of symmetric matrices $\mathbb{M}_n$. On the other hand, $x^{[2]}$ has linear dynamics as shown below. 
\begin{lemma}\label{lemma:dz}
    Along the trajectories of \eqref{sys:M}, one has 
    \begin{align}
        \frac{d}{dt}x^{[2]}=\bar Ax^{[2]}+\bar Ux.
    \end{align}
    with $\bar A:=D_n^\top(A\oplus A)(D_n^+)^\top$ and $\bar U:=D_n^\top(Bu\oplus Bu)$.
\end{lemma}

At first, it is tempting to consider the extended state $z^\top:=[(x^{[2]})^\top\;x^\top]$ which leads, as a consequence of Lemma \ref{lemma:dz}, to an LTV system of the form \eqref{sys:noC:ext}. The number of added auxiliary states is $n(n+1)/2$. However, we propose hereafter a procedure to extend the original system with {\it a minimum} number of auxiliary states to bring it to the LTV form. 

First, note that the vector space $\mathrm{span}\{C_1,\cdots,C_q\}$ has usually a lower dimension compared to $n(n+1)/2$. Let $\mathrm{dim}\;\mathrm{span}\{C_1,\cdots,C_q\}=\mathrm{rank}(\bar C)=:p_0$. Then, there exist distinct symmetric matrices $L_1,\cdots,L_{p_0}$ such that 
\begin{align}
    \mathrm{span}\{C_1,\cdots,C_q\}=\mathrm{span}\{L_1,\cdots,L_{p_0}\}.
\end{align}
It follows that the rows of $\bar Cx^{[2]}$ can be expressed as linear combinations of the quadratic forms $\frac{1}{2}x^\top L_1x,\cdots, \frac{1}{2}x^\top L_{p_0}x$. In fact, let us define the full row rank matrix $\bar L_0\in\mathbb{R}^{p_0\times n(n+1)/2}$ as 
\begin{align}
    \bar L_0:=\frac{1}{2}\begin{bmatrix}
    \vech(L_1)&
    \cdots&
    \vech(L_{p_0})
    \end{bmatrix}^\top .
\end{align}
Note that this matrix can be readily obtained from the rank factorization $\bar C=F\bar L_0$ for some full column rank matrix $F\in\mathbb{R}^{q\times p_0}$. Let us define the variable $\xi_0\in\mathbb{R}^{p_0}$ such that
\begin{equation}\label{eq:xi0}
    \xi_0:=\bar L_0x^{[2]}=\begin{bmatrix}
    \frac{1}{2}x^\top L_1x&
    \cdots&
     \frac{1}{2}x^\top L_{p_0}x 
    \end{bmatrix}^\top.
\end{equation}
Note that the output vector is linear in $\xi_0$ and $x$ since $y=F\xi_0+Dx$. As a result of Lemma \ref{lemma:dz}, the derivative of $\xi_0$ along the trajectories of \eqref{sys:M} satisfies
\begin{equation}\label{eq:dxi0}
  \begin{aligned}
    \dot\xi_0&=\bar L_0\bar Ax^{[2]}+\bar L_0\bar Ux.
\end{aligned}  
\end{equation}
Similar to \cite[Theorem 1]{montenbruck2017linear}, $\dot\xi_0$ will be linear in $\xi_0$ and $x$ if and only if $\mathrm{span}\{L_1,\cdots,L_{p_0}\}$ is $\mathbf{L}_A-$invariant or, equivalently, $\mathrm{ker}(\bar L_0)$ is $\bar A-$invariant. If this condition holds, the procedure stops by defining the extended state $z^\top:=[\xi_0^\top\;x^\top]$ which has linear dynamics. When this condition does not hold, we proceed as follows to add additional auxiliary states. Let us define $p_1\in\mathbb{N}$ such that
\begin{equation}\label{eq:p1}
    p_1:=\mathrm{rank}\begin{bmatrix}
      \bar L_0\\
      \bar L_0\bar A
    \end{bmatrix}-p_0.
\end{equation}
Note that $\mathrm{ker}(\bar L_0)$ is not $\bar A-$invariant which implies that $p_1\neq 0$. In this case, there exist a permutation matrix $P_0\in\mathbb{R}^{p_0\times p_0}$, a matrix $M_0^{(0)}\in\mathbb{R}^{(p_0-p_1)\times p_0}$ and a full row matrix $\bar L_1\in\mathbb{R}^{p_1\times n(n+1)/2}$ satisfying 
\begin{equation}\label{eq:P0L0A}
    P_0\bar L_0\bar A=\begin{bmatrix}
        \bar L_1\\
        M_0^{(0)}P_0\bar L_0
    \end{bmatrix}.
\end{equation}
Basically, $\bar L_1$ contains all rows of $\bar L_0\bar A$ that cannot be expressed as a linear combination of rows of $\bar L_0$. By defining $\xi_1:=\bar L_1x^{[2]}$, it follows from \eqref{eq:dxi0} and \eqref{eq:P0L0A} that
\begin{align}
     P_0\dot\xi_0&=\begin{bmatrix}
         0\\M_0^{(0)}
     \end{bmatrix}P_0\xi_0+\begin{bmatrix}
         I_{p_1}\\
         0
     \end{bmatrix}\xi_1+P_0\bar L_0\bar Ux.
\end{align}

\begin{algorithm}
\caption{Computation of the matrices $\bar L_k, P_k, M_i^{(k)}$}\label{algorithm1}
\begin{algorithmic}[1]
\Require Output matrix $\bar C$ and matrix $\bar A$
\State Compute $p_0:=\mathrm{rank}(\bar C)$
\State Rank factorize $\bar C=F\bar L_0$ with $\mathrm{rank}(\bar L_0)=p_0$
\For{$k\in\{1,2,\cdots\}$}
\State Calculate
\begin{equation}\label{eq:pk+1}
    p_k:=\mathrm{rank}\begin{bmatrix}
        \bar L_0\\
        \vdots\\
        \bar L_{k-1}\\
        \bar L_{k-1}\bar A
    \end{bmatrix}-\sum_{i=0}^{k-1} p_i
\end{equation}
\State Find a permutation matrix $P_{k-1}\in\mathbb{R}^{p_{k-1}\times p_{k-1}}$, matrices $M_i^{(k-1)}\in\mathbb{R}^{(p_{k-1}-p_k)\times p_i}$ and a full row matrix $\bar L_k\in\mathbb{R}^{p_k\times n(n+1)/2}$ satisfying\footnotemark
\begin{equation}\label{eq:PLA}
   P_{k-1} \bar L_{k-1}\bar A=\begin{bmatrix}
        \bar L_k\\
        \sum_{i=0}^{k-1}M_i^{(k-1)}P_i\bar L_i
    \end{bmatrix}
\end{equation}
\If{$p_{k}=0$}
\State Define $m=k$
\State \Return the matrices $\bar L_{k-1}, P_{k-1}, M_{i-1}^{(k-1)}$ for all $1\leq i\leq k\leq m$.
\EndIf
\EndFor
\end{algorithmic}
\end{algorithm}
\footnotetext{If $p_k=p_{k-1}$ (resp. $p_k=0$), matrices $M_i^{k-1}$ (resp. $\bar L_k$) are empty.}

The rank computation \eqref{eq:p1} and the decomposition procedure in \eqref{eq:P0L0A} is then iteratively repeated to obtain matrices  $\bar L_k, P_k, M_i^{(k)}$ and stops only when $p_k=0$. This is summarized in Algorithm \ref{algorithm1}. Note that, with $\xi_k:=\bar L_kx^{[2]}$ and since the rows of $\bar L_k$ are linearly independent (by construction), the execution of Algorithm \ref{algorithm1} necessary stops before the number of auxiliary states in $\xi_0,\cdots,\xi_{m-1}$ reaches $n(n+1)/2$. In fact, the additional states introduced by Algorithm \ref{algorithm1} is \begin{equation}
    \sum_{k=0}^{m-1} p_k=\mathrm{rank}\begin{bmatrix}
        \bar L_0\\
        \vdots\\
        \bar L_{m-1}
    \end{bmatrix}\leq n(n+1)/2.
\end{equation}
Our iterative procedure allows us to obtain, by construction, the minimum number of auxiliary states necessary to bring system \eqref{sys:M} to the LTV form \eqref{sys:noC:ext}. It follows from Lemma \ref{lemma:dz} and \eqref{eq:PLA} that
\begin{equation}
   \begin{aligned}
     P_k\dot\xi_k&=P_k\bar L_k\bar Ax^{[2]}+P_k\bar L_k\bar Ux\\
     &=\begin{bmatrix}
        \bar L_{k+1}\\
        \sum_{i=0}^kM_i^{(k)}P_i\bar L_i
    \end{bmatrix}x^{[2]}+P_k\bar L_k\bar Ux\\
    &=
     \sum_{i=0}^k\begin{bmatrix}
         0\\M_i^{(k)}
     \end{bmatrix}P_i\xi_i+\begin{bmatrix}
         P_{k+1}^\top\\
         0
     \end{bmatrix}P_{k+1}\xi_{k+1}+P_k\bar L_k\bar Ux,
\end{aligned} 
\end{equation}
where $k=0,\cdots,m-1$ with $\xi_{m}\equiv 0$. Now, if we define the \textit{extended state vector} 
\begin{equation}
z:=\begin{bmatrix}
    P_{m-1}\xi_{m-1}\\
    \vdots\\
    P_0\xi_0\\
    x
\end{bmatrix}:=\begin{bmatrix}
    P_{m-1}\bar L_{m-1}x^{[2]}\\
    \vdots\\
    P_0\bar L_0x^{[2]}\\
    x
\end{bmatrix} \in\mathbb{R}^{n+\sum_{k=0}^{m-1} p_k},
\end{equation}
the resulting dynamics are given by the LTV system \eqref{sys:noC:ext} with matrices
\begin{align}\label{eq:Au-3}
    \mathcal{A}(u)&=\begin{bmatrix}
   \begin{matrix}
       \bar M_{m-1}^{(m-1)}&\cdots&\cdots&\bar M_0^{(m-1)}
     \\\bar P_{m-1}&\ldots&\cdots&\bar M_0^{(m-2)}\\
        \vdots&\ddots&\vdots&\vdots\\
        0&\cdots&\bar P_1&M_0^{(0)}
    \end{matrix}&\rvline&\begin{matrix}
        P_{m-1}\bar L_{m-1}\bar U\\
        P_{m-2}\bar L_{m-2}\bar U\\
        \vdots\\
        P_{0}\bar L_{0}\bar U\\
    \end{matrix}
    \\\cmidrule(lr){1-3}
    0&\rvline&A
\end{bmatrix},\\
\mathcal{B}&=\left[\begin{matrix}
0\\\cmidrule(lr){1-1}B
\end{matrix}\right],\;
\mathcal{C}=\begin{bmatrix}\label{eq:C-3}
    0&\cdots&0&FP_0^\top&\rvline&D
\end{bmatrix},
\end{align}
where we defined
\begin{equation}\label{eq:M-P}
    \bar M_i^{(k)}:=\begin{bmatrix}
         0_{p_{k+1}\times p_i}\\M_i^{(k)}
     \end{bmatrix},\quad \bar P_k:=\begin{bmatrix}
         P_k^\top\\
         0_{(p_{k-1}-p_k)\times p_k}
     \end{bmatrix}.
\end{equation}
The following theorem summarizes the main result whose proof follows directly from the above derivations.
\begin{thm}
Consider system \eqref{sys:M}  along with matrices $A, B, C_q, d_q$ of appropriate dimensions. Suppose the system involves multiple outputs ($q>1$). Let $\bar L_k, P_k, M_i^{(k)}$ be the $k$-th matrices obtained from Algorithm \ref{algorithm1}, for $0\leq i\leq k\leq m-1$. Then, system \eqref{sys:M} can be embedded into the higher dimensional LTV system \eqref{sys:noC:ext} with matrices $\mathcal{A}(u)$, $\mathcal{B}$, and $\mathcal{C}$ given by \eqref{eq:Au-3}-\eqref{eq:M-P}.
\end{thm}
\begin{rem}
The procedure for the single-output case of Section \ref{ssec:Mot} is a particular case of the above-proposed procedure for the multiple-output case. In fact, for a single-output system, we have $\bar C=\bar L_0=\frac{1}{2}\vech(C)$ and $p_0=1$ (row vector). When $p_k=1$ in \eqref{eq:pk+1}, the matrices $M_i^{(k-1)}$ are empty, $P_{k-1}=1$, and the iteration in \eqref{eq:PLA} reduces to $\bar L_k=\bar L_{k-1}\bar A$. This iteration is equivalent to $\mathbf{L}_A^{[k]}(C)=\mathbf{L}_A\left(\mathbf{L}_A^{[k-1]}(C)\right)$  with $\vech(\mathbf{L}_A^{[k]}(C)):=\bar L_k$ (see equation \eqref{eq:vechQm} in the Appendix) . The LTV system's matrices in \eqref{eq:Au-3}-\eqref{eq:M-P} match those in \eqref{matrix:Au:ext}-\eqref{matrix:Cu:ext}.
\end{rem}
\begin{figure}[t!]
    \centering
    \includegraphics[width= 0.96\linewidth]{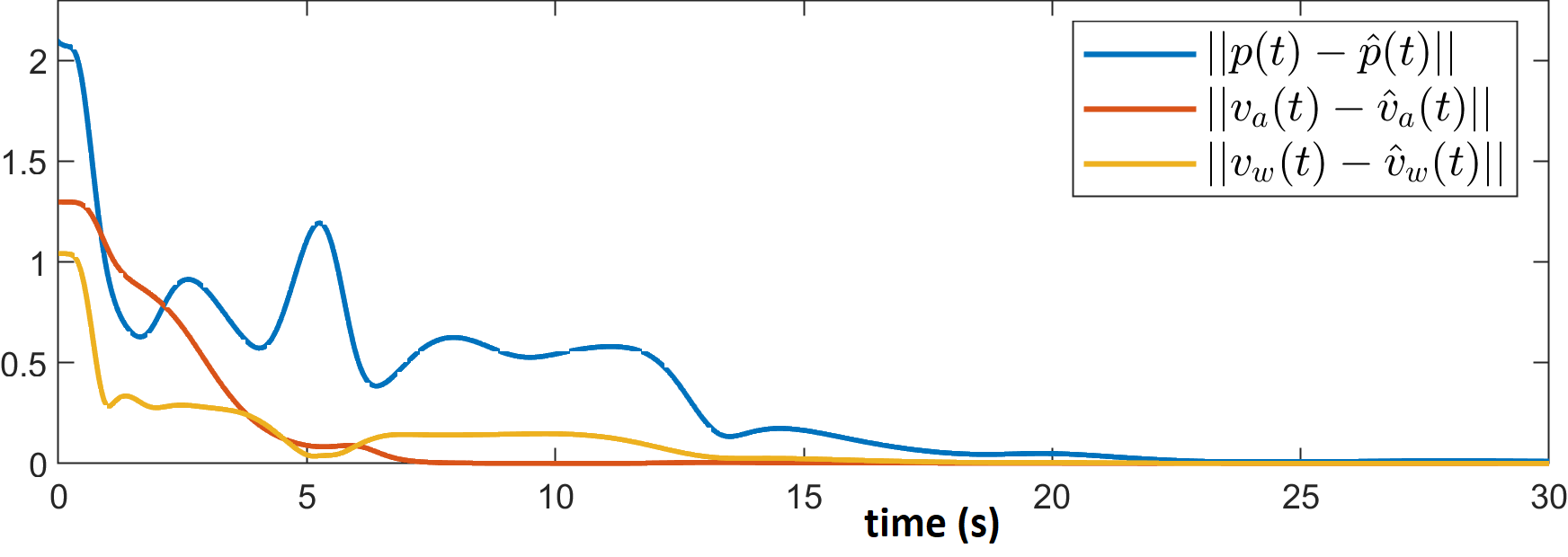}
    \caption{Convergence of the position error $||p(t)-\hat{p}(t)||$, air velocity error $||v_a(t)-\hat{v}_a(t)||$, and the wind velocity error $||v_w(t)-\hat{v}_w(t)||$.}
    \label{fig:error}
\end{figure}

\begin{rem}
The resulting extended system is a bilinear system with a linear output of the form \eqref{sys:noC:ext}. For certain input $u(t)$, this system can be considered as an LTV system. However, not every input $u(t)$ renders the system observable. The design of an observer to solve the state estimation problem follows by considering persistently exciting inputs  $u(\cdot)$ (inputs that render the system uniformly observable), see \cite{B07,Morin}. Under the uniform observability assumption, system \eqref{sys:noC:ext} admits the following Kalman-type observer:
\begin{align}\label{eq:dxhat_0}
	\dot{\hat z}=\mathcal{A}(u(t))\hat z+\mathcal{B}u(t)+P(t)\mathcal{C} ^\top Q(t)(y-\mathcal{C} \hat z)
\end{align}
such that $P(t)$ is the solution to the following continuous Riccati equation (CRE):
\begin{equation*} 
\begin{aligned}
\dot P(t)=&\mathcal{A}(u(t))P(t)+P(t)\mathcal{A}(u(t))^\top\\& -P(t)\mathcal{C}^\top Q(t)\mathcal{C}P(t)+V(t)
\end{aligned}
\end{equation*}
where $P(0)$ is positive definite and $Q(t)$ and $V(t)$ are continuously differentiable, uniformly bounded, and uniformly positive-definite matrix functions. Note that in a stochastic setting, the above estimator corresponds to the optimal Kalman filter where matrices $V(t)$ and $Q(t)^{-1}$ are interpreted as covariance matrices of additive noise on the system state and output \cite{Kalman1961new}. A variety of sufficient conditions for uniform observability can also be found in \cite{TBD20} and \cite{wang2021nonlinear}.
\end{rem}

\textit{Example 3:}\label{example4} Consider the system $\dot p=v_a+v_w$ with $\dot v_a=u$ and $\dot v_w=0$, where $p,v_a,v_w\in\mathbb{R}^n$ are, respectively, the position of a vehicle, the air velocity, and the wind velocity. The outputs are the range $y_1=0.5\|p\|^2$ and airspeed $y_2=0.5\|v_a\|^2$. Algorithm \ref{algorithm1} returns $m=3$ with $\xi_2=2\|v_a+v_w\|^2$, $\xi_1=2p^\top(v_a+v_w)$ and $\xi_0=[\|p\|^2\;\|v_a\|^2]$. One can easily verify that these additional states allow to bring this system to an LTV form for any $n\in\mathbb{N}$.   Fig. \ref{fig:error} shows the convergence of the error between the real states $p,v_a,v_w$ and the observed states $\hat{p},\hat{v}_a,\hat{v}_w$, with  $p(0)=(0,0,2)^\top$, $v_a(0)=(0,1,0)^\top$, $v_w(0)=(0,0,1)^\top$ and input $u(t)=(-\cos(t); -\sin(t); 0.5\cos(0.5t))^\top$, by using a Kalman-type observer (see \cite{B07}). UO analysis is left to the reader.

\section{Conclusion}\label{sec:conclusion}
We proposed an immersion-type technique that transforms linear time-invariant systems with quadratic outputs to a new linear time-varying system with linear output by adding a minimum finite number of auxiliary states to the original system. Both cases with single output and multiple outputs are considered. The state of the resulting LTV system can be globally estimated using Kalman-type observers provided the observability conditions necessary for the convergence of these observers are met. It is worth mentioning that the state matrix of the new LTV system depends explicitly on the input; therefore, this system's observability is tightly related to the \textit{richness} of the input signal. For instance, {\it uniform observability} of the resulting LTV system, given well-conditioned inputs (bounded and continuously differentiable), ensures that Riccati observers (refer to, for instance, \cite{HS17}) globally exponentially estimate the state of the system. An interesting future direction is to consider linear-time varying systems with quadratic outputs.

\appendix
\subsection{Proof of Lemma \ref{lemma:vech-LA}}
This identity is proved as follows: 
\begin{equation*}
\begin{aligned}
      &\vech(\mathbf{L}_A(X))\\
      &=D_n^+\vec(\mathbf{L}_A(X))=D_n^+\vec(XA)+D_n^+\vec(A^\top X)\\
      &\stackrel{\eqref{eq:KP1}}{=}D_n^+(A^\top\otimes I_n)\vec(X)+
       D_n^+(I_n\otimes A^\top)\vec(X)\\
      &\stackrel{\eqref{eq:KP6}}{=}D_n^+(A\oplus A)^\top\vec(X)=D_n^+(A\oplus A)^\top D_n\vech(X).
\end{aligned}
\end{equation*}
\subsection{Proof of Lemma \ref{lemma:Qm}}
Matrix $\bar A:=D_n^\top(A\oplus A)^\top (D_n^+)^\top$ satisfies: 
\begin{equation}\label{eq:vechQm}
\begin{aligned}
      \vech(\mathbf{L}_A^{[m]}(Q))&=\bar A^{m\top}\vech(\mathbf{L}_A^{[0]}(Q)),\quad\forall m\in\mathbb{N},
\end{aligned}
\end{equation}
in view of Lemma \ref{lemma:vech-LA}. To prove the existence of $m\leq n(n+1)/2$ satisfying the result of the lemma, it is sufficient to show that this result holds for $m=n(n+1)/2=:\bar n$. 
According to Cayley-Hamilton Theorem, the $n(n+1)/2\times n(n+1)/2$ square matrix $\bar A$, defined above, satisfies 
$\bar A^{\bar n}=\sum_{k=0}^{\bar n-1}\alpha_k\bar A^k,
$
for some coefficients $\alpha_0,\alpha_1,\cdots,\alpha_{\bar n-1}$. Therefore, one has
\begin{equation*}
\begin{aligned}
     &\vech(\mathbf{L}_A^{[\bar n]}(Q))= (\bar A^{\bar n})^\top\vech(Q)   =\sum_{k=0}^{\bar n-1}\alpha_k\bar A^{k\top}\vech(Q)\\
     &=\sum_{k=0}^{\bar n-1}\alpha_k\vech(\mathbf{L}_A^{[k]}(Q))  =\vech\left(\sum_{k=0}^{\bar n-1}\alpha_k\mathbf{L}_A^{[k]}(Q)\right).
\end{aligned}    
\end{equation*}
The last equation shows that $\mathbf{L}_A^{[\bar n]}\in\mathrm{span}\{\mathbf{L}_A^{[0]}(Q), \mathbf{L}_A^{[1]}(Q),\ldots,\mathbf{L}_A^{[\bar n-1]}(Q)\}$ and, hence, $\mathrm{span}\{\mathbf{L}_A^{[0]}(Q), \mathbf{L}_A^{[1]}(Q),\ldots,\mathbf{L}_A^{[\bar n-1]}(Q)\}$ is $\mathbf{L}_A-$invariant. The proof is complete.
\subsection{Proof of Lemma \ref{lemma:dz}}
In view of \eqref{sys:M}, one has
    \begin{equation*}
  \begin{aligned}
    \frac{d}{dt}&x^{[2]}=\frac{d}{dt}\left(D_n^\top(x\otimes x)\right)\\
    &=D_n^\top((Ax+Bu)\otimes x+x\otimes(Ax+Bu))\\
     &=D_n^\top((Ax)\otimes x+x\otimes(Ax)+(Bu)\otimes x+x\otimes(Bu))\\
     &\stackrel{\eqref{eq:KP3}-\eqref{eq:KP6}}{=}D_n^\top((A\oplus A)(x\otimes x)+(Bu\oplus Bu)x)\\
     &\stackrel{\eqref{eq:KP9}}{=}
    D_n^\top(A\oplus A)(D_n^+)^\top D_n^\top (x\otimes x)+D_n^\top(Bu\oplus Bu)x\\
    &=\bar AD_n^\top (x\otimes x)+\bar Ux=\bar Ax^{[2]}+\bar Ux.
\end{aligned}  
\end{equation*}

\bibliographystyle{IEEEtranS}
\bibliography{observer}

\end{document}